\documentclass[preprint,showpacs,preprintnumbers,amsmath,amssymb,superscriptaddress]{revtex4}

\usepackage{graphicx,graphics}   

\usepackage{graphicx,amsfonts}
\usepackage{epsfig}

\usepackage{dcolumn}
\usepackage{bm}
\begin{document}

\title{
	Angular momentum and parity of a two gluon system
}

\author{V. Pleitez}%
\email{vicente@ift.unesp.br}
\affiliation{
Instituto  de F\'\i sica Te\'orica--Universidade Estadual Paulista \\
R. Dr. Bento Teobaldo Ferraz 271, Barra Funda\\ S\~ao Paulo - SP, 01140-070,
Brazil
}

\date{01/19/2018}

\begin{abstract}
In this work we obtain all the allowed values of the angular momentum and parity for two massless non-Abelian fields. In particular, we show that, unlike the two photons system, two gluons can be in a state with $J^P=1^\pm$, $(2k)^-$, and $(2k+1)^-$, $k\geq1$. This is a general result since it was obtained using only the laws of addition of the angular momentum. 
\end{abstract}

\pacs{ 
4.70.Dj 	
12.38.-t 
14.70.-e 	
}     

\maketitle

\section{Introduction}
\label{sec:intro}
That a stationary system with spin-1 cannot decay into two photons was noted for the first time by Landau~\cite{landau} and, independently, by Yang~\cite{yang}. This is now known as the  Landau-Yang (LY) theorem. 
So far it has been assumed that the theorem should be valid also for non-Abelian theories such as QCD. Hence, the formation of spin-1 hadron was considered either by using off-shell gluons or three gluons, i.e., $gg\to \chi_{c,b}(1^+)$ is not allowed \textit{if} the LY theorem is valid~\cite{Shuvaev:2015fta}.  
However, although recently it has been shown that the spin-1 state is allowed for the two gluon system~\cite{Beenakker:2015mra,Cacciari:2015ela}, nothing was said about other odd values for the spin nor about the parity of these states. 

Here we will shown that the same arguments that forbid a two photon system to be in a spin-1 state, allow that two on-shell gluons to have this value of spin. Moreover,  we will obtain for the first time, in the context of a non-Abelian group, \textit{all} the allowed (and forbidden) values of the angular momentum and parity ($J^P$) for the system of two massless non-Abelian vector fields. These results are obtained using only the rules for addition of angular momentum and other well-known basic physics assumptions and, for this reason, our results do not depend on the model. Hence, although they are valid for a non-Abelian arbitrary group, we will have in mind $SU(n)$, in particular quantum chromodynamic (QCD) when $n=3$. The explicit and implicit hypotheses on which the Landau-Yang theorem is based were discussed in Ref.~\cite{Pleitez:2015cpa}.

The wave functions of the system formed by two non-Abelian massless vector fields are represented by tensors of $SU(n)$ with indices denoted by $a,b$, which are also  Lorentz second rank tensors in the spacial coordinates $i,j$. Hence, we denote these tensors as  $G^{ab}_{ik}$ and they include the angular momentum part written in terms of spherical harmonics, and  is  also bilinear in the components  of each vector field.

The  assumptions we will use are the following: Rotational invariance i.e., the conservation of the angular momentum, and the tensor $G^{ab}_{ik}$ is a function only of $\hat{n}$ defined as  $\vec{r}_2-\vec{r}_1=\hat{n}r$, and $r$ is fixed. Although the division of the total angular momentum into the spin and orbital angular momentum, $J=S+L$, has no physical meaning for massless particles, it can be used just as a mathematical trick~\cite{landau}. 
Moreover, since gluons are on-shell, they satisfy the transversality conditions:
\begin{equation}
G^{ab}_{ik}(\hat{n})n_k=0,\quad G^{ab}_{ik}(\hat{n})n_i=0.
\label{real1}
\end{equation}  
A two gluon system is a bosonic system, under the permutation of the color \textit{and} spacial indices( and also doing $\hat{n}\to -\hat{n}$), hence the total wave function must be symmetric,
\begin{equation}
G^{ab}_{ik}(\hat{n})=G^{ba}_{ki}(-\hat{n}).
\label{bose1}
\end{equation}

The tensor $G^{ab}_{ik}(\hat{n})$  can be written as the sum of two tensors, one being symmetric and the other antisymmetric in the color indices,
\begin{equation}
G^{ab}_{ik}(\hat{n})\,=G^{\{ab\}}_{ik}(\hat{n})+G^{[ab]}_{ik}(\hat{n}).
\label{soma1}
\end{equation}

The parity of each states is determined by how the respective tensor behaves under $\hat{n}\to -\hat{n}$, while
the spin of a two-gluons system is determined by the range of the Lorentz tensor (scalar, vector, second rank tensor,...), and the orbital angular momentum by the order of the spherical harmonics describing the states~\cite{landau}. 
Finally, we will work in the center of momentum frame of the gluons i.e., $\vec{p}_1+\vec{p}_2=0$.
Using the above assumptions we will obtain all the values, allowed and forbidden, of the parity and angular momentum of the two-gluons system. 

Below,  we will consider separately the two cases: i) when the system is symmetric in the color indices, $G^{\{ab\}}_{ik}(\hat{n})$ and, ii) when it is anti-symmetric, $G^{[ab]}_{ik}(\hat{n})$.

\section{Two-gluons in a symmetric color state}
\label{sec:2s}
 First let us consider the case when the tensor is symmetric in $a,b$, and the tensor is written as a sum of a symmetric and a anti-symmetric tensor in the Lorentz indices: 
\begin{equation}
G^{\{ab\}}_{ik}(\hat{n})=S^{\{ab\}}_{ik}(\hat{n})+ A^{\{ab\}}_{ik}(\hat{n}),
\label{soma2}
\end{equation}
and from (\ref{real1}) it follows 
\begin{equation}
G^{\{ab\}}_{ik}(\hat{n})n_k=0,\quad G^{\{ab\}}_{ik}(\hat{n})n_i=0.
\label{real2}
\end{equation}  
To be consistent with (\ref{bose1}), we have that $A^{\{ab\}}_{ik}(\hat{n})=-A^{\{ab\}}_{ik}(-\hat{n})$ and the states build from this tensor have odd parity, but $S^{\{ab\}}_{ik}(\hat{n})=S^{\{ab\}}_{ik}(-\hat{n})$, and now the respective states have positive parity.

Let us begin by obtaining the states that can be build from the anti-symmetric Lorentz tensor, $A^{\{ab\}}_{ik}(\hat{n})$. Since in three dimensions an antisymmetric tensor $A^{\{ab\}}_{ik}(\hat{n})$ is dual to a vector $A^{\{ab\}}_l(\hat{n})$ (or $\vec{A}^{\{ab\}}(\hat{n})$), we can write $A^{\{ab\}}_l(\hat{n})~=~(1/2)\varepsilon_{lik}G^{\{ab\}}_{ik}(\hat{n})$. Taking into account  the conditions in Eq.~(\ref{real1}) we have that $\vec{A}^{\{ab\}}\times \hat{n}=0$ or, that the vector $\vec{A}^{\{ab\}}$ is proportional to $\hat{n}$~:~$\vec{A}^{\{ab\}}(\hat{n})=\phi^{\{ab\}}(\hat{n})\hat{n}$, where $\phi^{\{ab\}}(\hat{n})$ is a Lorentz scalar function $\phi^{\{ab\}}(\hat{n})=\phi^{\{ab\}}(-\hat{n})$. The vector character of $A^{\{ab\}}_i(\hat{n})$ explains its odd parity and the true tensor character is given by the scalar $\phi^{\{ab\}}(\hat{n})$ that is, the states from $A^{\{ab\}}_{ik}(\hat{n})$ have spin $S=0$ and $J=L$ even.
Thus, from this tensor the only states allowed are those with 
\begin{equation}
J^P=0^-, (2k)^-,
\label{a1}
\end{equation}
while states with $J^P=1^-,(2k+1)^-$ are strictly for\-bi\-dden, here and below $k\geq1$.

On the other hand, the symmetric tensor $S^{\{ab\}}_{ik}(\hat{n})$ allows states with even parity and are written in terms of spherical harmonic of even order. This tensor can be decomposed as follows (the trace is only over the spacial indices): $S^{\{ab\}}_{ik}(\hat{n})~=~\textrm{Tr}\,S^{\{ab\}}_{ij}(\hat{n})~+~S^{\prime{\{ab\}}}_{ik}(\hat{n})$ with $\textrm{Tr}\ S^{\prime{\{ab\}}}_{ik}(\hat{n})~=~0$. 
The states from $\textrm{Tr}\,S^{\{ab\}}_{ij}(\hat{n})$ have spin $S=0$ and $J=L$ even: $J^P=0^+,(2k)^+$. From $S^{\prime{\{ab\}}}_{ik}(\hat{n})$ the states have spin $S=2$ and $L$ even. For even $J\not=0$ the satates allowed are three with $J-2,J,J+2$; for odd $J\not=1$ the satates have $J-1,J+1$. When $J=0,1$ the only state have $L=2$. Hence all
the states that can be built, in principle, from the symmetric tensor are:
\begin{eqnarray} 
&& J^P=0^+, (2k)^+, \; \textrm{from}\; \textrm{Tr}\,S^{\{ ab\}}_{ik},\nonumber \\ &&
nJ^P=0^+,1^+,3(2k)^+,2(2k+1)^+,\; \textrm{from}\; S^{\prime \{ab\}}_{ik}.
\label{s1}
\end{eqnarray}

Moreover, we have to take into account the conditions in Eq.~(\ref{real2}), i.e., the orthogonality between $S^{\{ab\}}_{ik}(\hat{n})$ and $\hat{n}$. It means that we have to exclude those states corresponding to a symmetric second rank tensor that is parallel to the vector $\hat{n}$, and which can be written as $s^{\{ab\}}_{ik}(\hat{n})=B^{\{ab\}}_i(\hat{n})n_k+B^{\{ab\}}_k(\hat{n})n_i$, with $B^{\{ab\}}_i(\hat{n})$ being a vector (it changes sign under $\hat{n}\to-\hat{n}$) has $S=1$ and odd $L$, with the states of even parity. This means that for even $J\not=0$ we have two states with $J-1,J+1$; for odd $J$ we have $J=L$. When $J=0$ we have only $L=1$. 
Hence, the states allowed by $s^{\{ab\}}_{ik}$ are 
\begin{equation} 
nJ^P=0^+,1^+,2(2k)^+,(2k+1)^+.
\label{s2}
\end{equation}
After subtracting from the states in Eq.~(\ref{s1}) the states in (\ref{s2}), we see that  only states with
\begin{equation}
J^P=0^+,2(2k)^+,(2k+1)^+,
\label{s3}
\end{equation}
can be built from the symmetric tensor, $S^{{\{ab\}}}_{ik}(\hat{n})$. The states $1^-$, $1^+$, and $(2k+1)^-$ are completely forbidden for the two-gluon system when the system is symmetric in the color indices.  These results are summarized in the second line of the Table. In the first line of the Table we show the angular momentum and parity allowed for the states of two-photon, obtained by the same arguments, and we see that these coincide with those of the two-gluon system when the latter system is symmetric in the color indices. 

Notice that these arguments has nothing to do with the existence, or not, of a direct interactions of the two gluons with a decaying system. In fact, in the well known cases of an scalar $J^P=0^+$ and a pseudo-scalar $J^P=0^-$, the connection with the two photon system is through charged fermion loop effects:  the Higgs boson $h\to \gamma\gamma$, and the pion $\pi^0\to\gamma\gamma$, respectively. 

\section{Two-gluons in an anti-symmetric color state}
\label{sec:2a}
Now, let us consider the case when the two-gluon system is anti-symmetric in the color degrees of freedom.
As in the previous case the wave functions do not change sign when the gluons are permuted in all their indices, and also $\hat{n}\to -\hat{n}$:
\begin{equation} 
G^{[ab]}_{ik}(\hat{n})=G^{[ba]}_{ki}(-\hat{n}).
\label{bose2}
\end{equation}
However, since the values of the angular momentum of the system depend only on the spacial indices, it is ne\-ce\-ssa\-ry to take  into account only the spacial indices, then
\begin{equation} 
G^{[ab]}_{ik}(\hat{n})=-G^{[ab]}_{ki}(-\hat{n}),
\label{bose3}
\end{equation}
and the system also must satisfy the transversality conditions, from Eq.~(\ref{real1}):
\begin{equation}
G^{[ab]}_{ik}(\hat{n})n_k=0,\quad G^{[ab]}_{ik}(\hat{n})n_i=0.
\label{real3}
\end{equation}
The tensor $G^{[ab]}_{ik}$ can be written as a sum of a symemtric and anti-symmetric tensor in the indices $i,k$,
\begin{equation}
G^{[ab]}_{ik}(\hat{n})=S^{[ab]}_{ik}(\hat{n})+A^{[ab]}_{ik}(\hat{n}).
\label{soma3}
\end{equation}
and the transversality conditions in Eq.~(\ref{real3}).

In this case $A^{[ab]}_{ik}(\vec{n})=+A^{[ab]}_{ik}(-\vec{n})$ and $S^{[ab]}_{ik}(\vec{n})=-S^{[ab]}_{ik}(-\vec{n})$. 
Thus, under a parity transformation, the states related to $S^{[ab]}_{ik}(\hat{n})$ are odd,  and those related to $A^{[ab]}_{ik}(\hat{n})$ are even. 

Now, the dual of the tensor $A^{[ab]}_{ik}(\hat{n})$ is 
a pseudo-vector,~$A^{[ab]}_l(\hat{n})\!\!=\!\!+A^{[ab]}_l(-\hat{n})$. 
By similar arguments to those in the previous case, we can show that this tensor is written 
as~$\vec{A}^{[ab]}(\hat{n})~=~\phi^{[ab]}_P(\hat{n})\hat{n}$, where  $\phi^{[ab]}_P(\hat{n})$ is now a Lorentz pseudo-scalar function i.e.,~$\phi^{[ab]}_P(\hat{n})=-\phi^{[ab]}_P(-\hat{n})$. The pseudo-vector character of $A^{[ab]}_i(\hat{n})$ explains its even parity and the true tensor character is given by the pseudo-scalar $\phi^{\{ab\}}_P(\hat{n})$ that is, the states from $A^{\{ab\}}_{ik}(\hat{n})$ have spin $S=0$ and $J=L$ odd. Then, the wave functions allowed from $A^{[ab]}_{ik}(\vec{n})$ have 
\begin{equation}
J^P=1^+,(2k+1)^+,
\label{a2}
\end{equation}
while $J^P=0^\pm,1^-,(2k+1)^-$ are completely forbidden from this tensor.

Next, let us consider the symmetric tensor $S^{[ab]}_{ik}(\hat{n})$ whose states have now odd parity, as can be seen from Eq.~(\ref{bose3}), thus they may be expanded in terms of spherical harmonics of odd order $L$. 
As before, we use the decomposition $S^{[ab]}_{ik}(\hat{n})=\textrm{Tr}\,S^{[ab]}_{ik}(\hat{n})+S^{\prime[ab]}_{ik}(\hat{n})$ with $\textrm{Tr}S^{\prime[ab]}_{ik}(\hat{n})=0$. The states from $\textrm{Tr}\,S^{[ab]}_{ik}(\hat{n})$ have $S=0$ and $J=L$  then $J$ is odd and the allowed wave functions can have $J^P=1^-,(2k+1)^-$. 
The states from the traceless $\mathcal{S}^{\prime[ab]}_{ik}(\hat{n})$ correspond to spin $S=2$ 
and odd $L$. We have that: i) for even $J\not=0$ there are two possible states $J-1,J+1$ i.e., 
$nJ^P=2(2k)^-$, and for odd $J\not=1$ we have $J-2,J,J+2$, i.e., $nJ^P=3(2k+1)^-$. Finally, for $J=0,1$, $L=2$ i.e., $J^P=0^-,1^-$. Thus, the wave functions up to this moment can have:
\begin{eqnarray} 
&& J^P=1^-,(2k+1)^-\; \mathrm{from}\; \mathrm{Tr}\,S^{[ab]}_{ik},\nonumber \\&&
nJ^P=0^-,1^-,2(2k)^-,3(2k+1)^-\;\mathrm{from}\; S^{\prime[ab]}_{ik}.
\label{a3}
\end{eqnarray} 
However, because of the conditions in Eq.~(\ref{real3}), we have to exclude from $S^{\prime[ab]}_{ik}$, the states obtained from the Lorentz second rank pseudo-tensor that are parallel to $\hat{n}$ and having odd parity. They are of the form $\textrm{s}^{[ab]}_{ik}(\hat{n})=V^{[ab]}_i(\hat{n})n_k+V^{[ab]}_k(\hat{n})n_i$, where $V^{[ab]}_i(\hat{n})$ is a axial-vector, $V^{[ab]}_i(\vec{n})=+V^{[ab]}_i(-\vec{n})$ and $\textrm{s}^{[ab]}_{ik}(\hat{n})=-\textrm{s}^{[ab]}_{ik}(-\hat{n})$. The axial-vector must be expressible in terms of spherical harmonics of even order $L$. The respective states have $S=1$ with odd parity. For even $J$  we have $J=L$ i.e., $J^P=0^-,(2k)^-$. For odd $J\not=1$, there are two states  $J-1,J+1$, i.e., $J^P=2(2k+1)^-$, finally for $J=1$ we have one state with $L=0$, $J^P=1^-$. Hence, the states allowed by $s^{[ab]}_{ik}$ are
\begin{equation}
nJ^P=0^-,1^-,(2k)^-,2(2k+1)^-.
\label{a4}
\end{equation}
After subtracting the states in Eq.~(\ref{a4}) from those in Eq.~(\ref{a3}),  we have that the states allowed for the anti-symmetric tensor $S^{[ab]}_{ik}(\hat{n})$ are only those with
\begin{equation}
nJ^P=1^-,(2k)^-,2(2k+1)^-,
\label{a5}
\end{equation}
We see that in this case, a massive particle with $J^P=1^-$ can decay into two gluons. Note also that the states $(2k+1)^\pm,(2k+1)^+$ are completely forbidden. See the third line in the Table.

\begin{table}[ht]
	\begin{tabular}{|l||c| }\hline  
		& $J^P$   \\ \hline \hline
		$\gamma\gamma$ & $0^-,(2k)^-$; $0^+$, 2$(2k)^+$,$(2k+1)^+$  \\ \hline
		$(gg)_s$ &  $0^-,(2k)^-$; $0^+$, 2$(2k)^+$,$(2k+1)^+$  \\ \hline
		$(gg)_a$ & $1^\pm$, $(2k)^-$, $2(2k+1)^-$ \\ \hline
	\end{tabular}
	\caption{Allowed angular momenta for two massless vector fields. Here $k\ge1$, see the notation in the text. $\gamma$ denotes a photon, and $g$ a gluon. $(gg)_s$ and $(gg)_a$ denotes symmetric and anti-symmetric, respectively, color degrees of freedom in the two-gluon system. All values of $J^P$ that do not appear in this table are strictily forbidden.
	}
\end{table}

\section{Conclusions}
\label{sec:con}

Here we have shown all the allowed values for the angular momentum and parity of a pair of on-shell gluons.   
These values include states with  $J^P=1^\pm,(2k+1)^-$ which are forbidden for the two photon system. 
We stress that our results are general since  they rest only on fundamental assumptions, namely  the laws of addition of the angular momentum and, for this reason, they are independent of the model and of the way the colored system with spin-1 is formed i.e., if it happens at a  given order of perturbation theory or even non-perturbatively. 

In fact, in Ref.~\cite{Ma:2014oha} it was noted that the case of $gg\to (q\bar{q})^8_{J=1}$ occurs at the 1-loop order. At first sight, this can be considered as an evasion of the Landau-Yang theorem.
However,  the same arguments which forbid that a two photon system have spin-1, allows that a two (on-shell) gluon system can be in a spin-1 statemkh, if the color degrees of freedom of the system are antisymmetric. Therefore, the process $ gg \to (q \bar {q}) ^ 8_ {J = 1} $ has a contribution in which both gluons are on-shell.
The fact that two gluons can be in a state with spin-1 was noted in Refs.~\cite{Beenakker:2015mra}, and in Ref.~\cite{Cacciari:2015ela} it was found that the amplitude of this decay cannot vanishes to all orders and the authors explain this result in terms of the emergence of higher-dimension operators, radiatively generated in the effective action already at one-loop order. 
The important point of our work is that now \textit{we know} what  values of the angular momentum and parity are allowed for these systems. Notice that only the case $J^P=(2k)^-$ is allowed for any two massless vector boson system. 

\vskip .3cm
\section*{Acknowledgments}
The author would like to thanks to CNPq for partial support and to FAPESP for the support under funding Grant No. 2014/19164-6.

\vskip 1.0cm

\end{document}